\documentclass[10pt,journal,compsoc]{IEEEtran}
\ifCLASSOPTIONcompsoc
  \usepackage[nocompress]{cite}
\else
  \usepackage{cite}
\fi
\ifCLASSINFOpdf
\else
\fi
\usepackage{amsmath}
\usepackage{siunitx}
\usepackage{graphicx}
\usepackage{optidef}
\usepackage{comment}
\usepackage{amssymb}
\usepackage{amsthm}
\usepackage{algorithm}
\usepackage{color}
\usepackage[noend]{algorithmic}
\newtheorem{definition}{definition}
\newtheorem{theorem}{theorem}
\newtheorem{lemma}{lemma}

\newcommand{\GeoGI}{road-network-indistinguishability}
\newcommand{\GeoI}{geo-indistinguishability}
\newcommand{\privacygain}{AE}

\newcommand{\roadi}{geo-graph-indistinguishability}
\newcommand{\rni}{GG-I}

\newcommand{\plm}{PLM}
\newcommand{\plmg}{PLMG}
\newcommand{\tp}{TP}
\newcommand{\gem}{GEM}
\newcommand{\fig}{Fig.}
\newcommand{\sect}{Section}
\newcommand{\mc}{PC}

\begin{document}
%
% paper title
% Titles are generally capitalized except for words such as a, an, and, as,
% at, but, by, for, in, nor, of, on, or, the, to and up, which are usually
% not capitalized unless they are the first or last word of the title.
% Linebreaks \\ can be used within to get better formatting as desired.
% Do not put math or special symbols in the title.
\title{Geo-Graph-Indistinguishability: Location Privacy on Road Networks Based on Differential Privacy}
%
%
% author names and IEEE memberships
% note positions of commas and nonbreaking spaces ( ~ ) LaTeX will not break
% a structure at a ~ so this keeps an author's name from being broken across
% two lines.
% use \thanks{} to gain access to the first footnote area
% a separate \thanks must be used for each paragraph as LaTeX2e's \thanks
% was not built to handle multiple paragraphs
%
%
%\IEEEcompsocitemizethanks is a special \thanks that produces the bulleted
% lists the Computer Society journals use for "first footnote" author
% affiliations. Use \IEEEcompsocthanksitem which works much like \item
% for each affiliation group. When not in compsoc mode,
% \IEEEcompsocitemizethanks becomes like \thanks and
% \IEEEcompsocthanksitem becomes a line break with idention. This
% facilitates dual compilation, although admittedly the differences in the
% desired content of \author between the different types of papers makes a
% one-size-fits-all approach a daunting prospect. For instance, compsoc 
% journal papers have the author affiliations above the "Manuscript
% received ..."  text while in non-compsoc journals this is reversed. Sigh.

\author{Shun~Takagi\textsuperscript{1}*\thanks{*corresponding author, E-mail: takagi.shun.45a@st.kyoto-u.ac.jp, full postal address: 36-1 Yoshida Honmachi, Sakyo-ku, Kyoto 606-8501 Japan},
        Yang~Cao\textsuperscript{1},
        Yasuhito~Asano\textsuperscript{2},
        Masatoshi~Yoshikawa\textsuperscript{1}\\% <-this % stops a space
%\IEEEcompsocitemizethanks{\IEEEcompsocthanksitem M. Shell was with the Department
%of Electrical and Computer Engineering, Georgia Institute of Technology, Atlanta,
%GA, 30332.\protect\\
% note need leading \protect in front of \\ to get a newline within \thanks as
% \\ is fragile and will error, could use \hfil\break instead.
%E-mail: see http://www.michaelshell.org/contact.html
%\IEEEcompsocthanksitem J. Doe and J. Doe are with Anonymous University.}% <-this % stops an unwanted space
\small{
\textsuperscript{1}\textit{Kyoto University}, 36-1 Yoshida Honmachi, Sakyo-ku, Kyoto 606-8501 Japan\\ 
\textsuperscript{2}\textit{Toyo University}, 1-7-11 Akabanedai, Kita-ku, Tokyo 115-0053 Japan}
\thanks{A shorter version of this paper appeared at the 33rd Annual IFIP WG 11.3
Conference on Data and Applications Security and Privacy (DBSec'19). This is the full version, published in.}}

% note the % following the last \IEEEmembership and also \thanks - 
% these prevent an unwanted space from occurring between the last author name
% and the end of the author line. i.e., if you had this:
% 
% \author{....lastname \thanks{...} \thanks{...} }
%                     ^------------^------------^----Do not want these spaces!
%
% a space would be appended to the last name and could cause every name on that
% line to be shifted left slightly. This is one of those "LaTeX things". For
% instance, "\textbf{A} \textbf{B}" will typeset as "A B" not "AB". To get
% "AB" then you have to do: "\textbf{A}\textbf{B}"
% \thanks is no different in this regard, so shield the last } of each \thanks
% that ends a line with a % and do not let a space in before the next \thanks.
% Spaces after \IEEEmembership other than the last one are OK (and needed) as
% you are supposed to have spaces between the names. For what it is worth,
% this is a minor point as most people would not even notice if the said evil
% space somehow managed to creep in.

% The paper headers
\markboth{Journal of \LaTeX\ Class Files,~Vol.~14, No.~8, August~2015}%
{Shell \MakeLowercase{\textit{et al.}}: Bare Demo of IEEEtran.cls for Computer Society Journals}
% The only time the second header will appear is for the odd numbered pages
% after the title page when using the twoside option.
% 
% *** Note that you probably will NOT want to include the author's ***
% *** name in the headers of peer review papers.                   ***
% You can use \ifCLASSOPTIONpeerreview for conditional compilation here if
% you desire.

% The publisher's ID mark at the bottom of the page is less important with
% Computer Society journal papers as those publications place the marks
% outside of the main text columns and, therefore, unlike regular IEEE
% journals, the available text space is not reduced by their presence.
% If you want to put a publisher's ID mark on the page you can do it like
% this:
%\IEEEpubid{0000--0000/00\$00.00~\copyright~2015 IEEE}
% or like this to get the Computer Society new two part style.
%\IEEEpubid{\makebox[\columnwidth]{\hfill 0000--0000/00/\$00.00~\copyright~2015 IEEE}%
%\hspace{\columnsep}\makebox[\columnwidth]{Published by the IEEE Computer Society\hfill}}
% Remember, if you use this you must call \IEEEpubidadjcol in the second
% column for its text to clear the IEEEpubid mark (Computer Society jorunal
% papers don't need this extra clearance.)

% use for special paper notices
%\IEEEspecialpapernotice{(Invited Paper)}

% for Computer Society papers, we must declare the abstract and index terms
% PRIOR to the title within the \IEEEtitleabstractindextext IEEEtran
% command as these need to go into the title area created by \maketitle.
% As a general rule, do not put math, special symbols or citations
% in the abstract or keywords.
\IEEEtitleabstractindextext{%

\begin{abstract}
In recent years, concerns about location privacy are increasing with the spread of location-based services (LBSs). Many methods to protect location privacy have been proposed in the past decades. Especially, perturbation methods based on Geo-Indistinguishability (Geo-I), which randomly perturb a true location to a pseudolocation, are getting attention due to its strong privacy guarantee inherited from differential privacy.
However, Geo-I is based on the Euclidean plane even though many LBSs are based on road networks (e.g. ride-sharing services). This causes unnecessary noise and thus an insufficient tradeoff between utility and privacy for LBSs on road networks. To address this issue, we propose a new privacy notion, Geo-Graph-Indistinguishability (GG-I), for locations on a road network to achieve a better tradeoff. We propose Graph-Exponential Mechanism (GEM), which satisfies GG-I. Moreover, we formalize the optimization problem to find the optimal GEM in terms of the tradeoff. However, the computational complexity of a naive method to find the optimal solution is prohibitive, so we propose a greedy algorithm to find an approximate solution in an acceptable amount of time. Finally, our experiments show that our proposed mechanism outperforms a Geo-I's mechanism with respect to the tradeoff.
\end{abstract}

% Note that keywords are not normally used for peerreview papers.
}

% make the title area
\maketitle

% To allow for easy dual compilation without having to reenter the
% abstract/keywords data, the \IEEEtitleabstractindextext text will
% not be used in maketitle, but will appear (i.e., to be "transported")
% here as \IEEEdisplaynontitleabstractindextext when the compsoc 
% or transmag modes are not selected <OR> if conference mode is selected 
% - because all conference papers position the abstract like regular
% papers do.
\IEEEdisplaynontitleabstractindextext
% \IEEEdisplaynontitleabstractindextext has no effect when using
% compsoc or transmag under a non-conference mode.

% For peer review papers, you can put extra information on the cover
% page as needed:
% \ifCLASSOPTIONpeerreview
% \begin{center} \bfseries EDICS Category: 3-BBND \end{center}
% \fi
%
% For peerreview papers, this IEEEtran command inserts a page break and
% creates the second title. It will be ignored for other modes.
\IEEEpeerreviewmaketitle

% Computer Society journal (but not conference!) papers do something unusual
% with the very first section heading (almost always called "Introduction").
% They place it ABOVE the main text! IEEEtran.cls does not automatically do
% this for you, but you can achieve this effect with the provided
% \IEEEraisesectionheading{} command. Note the need to keep any \label that
% is to refer to the section immediately after \section in the above as
% \IEEEraisesectionheading puts \section within a raised box.

% The very first letter is a 2 line initial drop letter followed
% by the rest of the first word in caps (small caps for compsoc).
% 
% form to use if the first word consists of a single letter:
% \IEEEPARstart{A}{demo} file is ....
% 
% form to use if you need the single drop letter followed by
% normal text (unknown if ever used by the IEEE):
% \IEEEPARstart{A}{}demo file is ....
% 
% Some journals put the first two words in caps:
% \IEEEPARstart{T}{his demo} file is ....
% 
% Here we have the typical use of a "T" for an initial drop letter
% and "HIS" in caps to complete the first word.

% You must have at least 2 lines in the paragraph with the drop letter
% (should never be an issue)

\IEEEraisesectionheading{\section{Introduction}\label{sec:introduction}}
\IEEEPARstart{I}{n} recent years, the spread of smartphones and GPS improvements have led to a growing use of location-based services (LBSs). While such services have provided enormous benefits for individuals and society, their exposure of the users' location raises privacy issues. Using the location information, it is easy to obtain sensitive personal information, such as information pertaining to home and family. In response, many methods have been proposed in the past decade to protect location privacy. These methods involve three main approaches: perturbation, cloaking, and anonymization. Most of these privacy protection methods are based on the Euclidean plane rather than on road networks; however many LBSs such as UBER\footnote{https://marketplace.uber.com/matching} and Waze\footnote{https://www.waze.com/ja/} are based on road networks to capitalize on their structures~\cite{cho2005efficient,kolahdouzan2004voronoi,papadias2003query}, resulting in utility loss and privacy leakage. Some prior works have revealed this fact~\cite{wang2009privacy, hossain2011h, duckham2005formal} and proposed methods that use road networks and are based on cloaking and anonymization. However, cloaking and anonymization also have weaknesses: if an adversary has peripheral knowledge about a true location, such as the range of a user's location, no privacy protection is guaranteed (in detail, we refer to \sect~\ref{sec:related}). In this paper, based on differential privacy~\cite{dwork2011differential}, we consider a perturbation method that does not possess such weakness.
First, we review perturbation methods and differential privacy~\cite{dwork2011differential}, which are the bases of our work; then, we describe the details of our work.

Perturbation methods modify a true location to another location by adding random noise~\cite{geo-i,shokri-strategy} using a mechanism. Shokri et al.~\cite{shokri-quantify} defined location privacy introduced by a mechanism, and they constructed a mechanism that maximizes location privacy. This concept of location privacy assumes an adversary with some knowledge; this approach cannot guarantee privacy against other adversaries. 

Differential privacy~\cite{dwork2011differential} has received attention as a rigorous privacy notion that guarantees privacy protection against any adversary. Andr$\acute{\rm e}$s et al.~\cite{geo-i} defined a formal notion of location privacy called geo-indistinguishability (Geo-I) by extending differential privacy. A mechanism that achieves it guarantees the indistinguishability of a true location from other locations to some extent against any adversary. 
However, because this method is based on the Euclidean plane, Geo-I does not tightly protect the privacy of locations on road networks, which results in a loose tradeoff between utility and privacy. In other words, Geo-I protects privacy too much for people on road networks.

%%% takagi 12/19
Geo-I assumes only that the given data is a location, which causes a loose tradeoff between utility and privacy for LBSs over road networks.
We make an assumption that a user is located on a road network. We model the road network using a graph and following this assumption, we propose a new privacy definition, called $\epsilon$-geo-graph-indistinguishability (GG-I), based on the notion of differential privacy. Additionally, we propose the graph-exponential mechanism (GEM), which satisfies GG-I.

Although GEM outputs a vertex of a graph that represents a road network, the output range (i.e., set of vertices) is adjustable, which induces the idea that there exists an optimal output range. Next, we introduce Shokri's notion~\cite{shokri-quantify} of privacy and utility, which we call adversarial error (AE) and quality loss (Q$_{loss}$), and analyze the relationship between output range and Shokri's notion. Moreover, we formalize the optimization problem to search the optimal range for AE and Q$_{loss}$. However, the number of combination of output ranges is $2^{|V|}$, where $|V|$ denotes the size of vertices, which makes it difficult to solve the optimization problem in acceptable time. Consequently, we propose a greedy algorithm to find an approximate solution to the optimization problem in an acceptable amount of time.

%%% takagi 12/19
Because our definition tightly considers location privacy on road networks, it results in a better tradeoff between utility and privacy. To demonstrate this aspect, we compare GEM with the baseline mechanism proposed in \cite{geo-i}. 
In our experiments on two real-world maps, GEM outperforms the baseline w.r.t. the tradeoff between utility and privacy. Moreover, we obtained the prior distribution of a user using a real-world dataset. Then, we show that the privacy protection level of a user who follows the prior distribution can be effectively improved by the optimization.
%%% end

In summary, our contributions are as follows:
\begin{itemize}
  \item We propose a privacy definition for locations on road networks, called $\epsilon$-geo-graph-indistinguishability (GG-I).
  \item We propose a graph-exponential mechanism (GEM) that satisfies GG-I.
  \item We analyze the performance of GEM and formalize optimization problems to improve utility and privacy protection.
  \item We experimentally show that our proposed mechanism outperforms the mechanism proposed in~\cite{geo-i} w.r.t. the tradeoff between utility and privacy and provide an optimization technique that effectively improves it.
\end{itemize}

\begin{comment}
Our contributions are threefold.
\begin{itemize}
  \item We propose a privacy definition on road networks, \roadi, and privacy gain and utility loss of its mechanism.
  \item We make a mechanism satisfying \GeoGI.
  \item We empirically show that our method outperforms a \GeoI method.
  \item We propose a criteria that indicates the mechanism performance for a given graph and a given prior distribution and the way to improve the mechanism w.r.t this indicator.
\end{itemize}
\end{comment}

% needed in second column of first page if using \IEEEpubid
%\IEEEpubidadjcol

\section{Preliminaries and Problem Setting}
In this section, we first review the formulations for a perturbation mechanism, empirical privacy gain and utility loss. Next, we describe the concept of differential privacy~\cite{dwork2011differential}, which is the basis of our proposed privacy notion. Finally, we explain a setting where we define privacy.

\subsection{Perturbation Mechanism on the Euclidean Plane}
Here, we explain the formulations for a perturbation mechanism, empirical privacy gain and utility loss~\cite{shokri-strategy}.
\subsubsection{User and Adversary}
\label{subsubsec:userandadv}
Shokri et al.~\cite{shokri-strategy} assumed that user $u$ is located at location $x\in{\mathbb{R}^2}$ according to a prior distribution $\pi_u(x)$. 
LBSs are used by people who wants to protect their location privacy but receive high-quality services. The user adopts a perturbation mechanism $M:\mathbb{R}^2\to\mathcal{Z}$ that sends a pseudolocation $M(x)=z\in\mathcal{Z}$ instead of his/her true location $x$ where $\mathcal{Z}\subseteq\mathbb{R}^2$. Assume that an adversary $a$ has some knowledge represented as a prior distribution about the user location $\pi_a(x)$ and tries to infer the user's true location from the observed pseudolocation $z$. In this paper, we assume that the adversary has unbounded computational power and precise prior knowledge, i.e., $\pi_a(x)=\pi_u(x)$. Although this assumption is advantageous for the adversary, protection against such an adversary confers a strong guarantee of privacy.

\subsubsection{Empirical Privacy Gain and Utility Loss}
\label{subsubsec:aeandsql}
The empirical privacy gain obtained by mechanism $M$ is defined as follows, which we call adversarial error (\privacygain).

\begin{equation}
\label{equ:privacygain}
\begin{split}
\nonumber
&\privacygain(\pi_a,M,h,d_q) =\\ &\sum_{\hat{x},x,z}\pi_a(x)\Pr(M(x)=z)\Pr(h(z)=\hat{x})d_q(\hat{x},x)
\end{split}
\end{equation}
\noindent
where $d_q$ is a distance over $\mathbb{R}^2$ and $h$ is a probability distribution over $\mathbb{R}^2$ that represents the inference of the adversary about the user's location. Thus, intuitively, \privacygain\ represents the expected distance between the user's true location $x$ and the location $\hat{x}$ inferred by the adversary.
Next, we explain the model of an adversary, that is, how an adversary constructs a mechanism $h$, which is called an optimal inference attack~\cite{shokri-strategy}. An adversary who obtains a user's perturbed location $z$ tries to infer the user's true location through an optimal inference attack. In this type of attack, the adversary solves the following mathematical optimization problem to obtain the optimal probability distribution and constructs the optimal inference mechanism $h$.
Then, by applying this mechanism to the input $z$, the adversary can estimate the user's true location.
\begin{mini}|l|
{h}{\privacygain(\pi_a,M,h,d_q)}{}{}
\addConstraint{\sum_{\hat{x}}\Pr(h(z)=\hat{x})}{=1}{, \forall z}
\addConstraint{\Pr(h(z)=\hat{x})}{\geq 0}{, \forall z,\hat{x}}
\nonumber
\end{mini}
For example, if an adversary knows a road network, the domain of his prior $\pi_a$ consists of locations on that road network, and $d_p$ is the shortest distance on the road network.
In this setting, the problem is a linear programming problem because $\Pr(h(z)=\hat{x})$ represents a variable and the other terms are constant; thus, the objective function and the constraints are linear. We solve this problem using CBC (coin-or branch and cut)\footnote{https://projects.coin-or.org/Cbc} solver from the Python PuLP library.

The utility loss caused by mechanism $M$, called quality loss (Q$^{loss}$), is defined as follows:
\begin{equation}\label{SQL_def}
\nonumber
    Q^{loss}(\pi_u,M,d_q) = \sum_{x,x^\prime}\pi_u(x) \Pr(M(x)=x^\prime)d_q(x,x^\prime)
\end{equation}
Q$^{loss}$ denotes the expected distance between the user's true location $x$ and the pseudolocation $z$.

\subsection{Differential Privacy}
\label{subsec:differentialprivacy}
Differential privacy~\cite{dwork2011differential} is a mathematical definition of the privacy properties of individuals in a statistical dataset. Differential privacy has become a standard privacy definition and is widely accepted as the foundation of a mechanism that provides strong privacy protection. $d\in\mathcal{D}$ denotes a record belonging to an individual and dataset $X$ is a set of $n$ records. When neighboring datasets are defined as two datasets which differ by only a single record, then $\epsilon$-differential privacy is defined as follows.
\begin{definition}[$\epsilon$-differential privacy]
Given algorithm $M:\mathcal{D}\to \mathcal{S}$ and the neighboring datasets $X, X^\prime \in \mathcal{D}$, the privacy loss is defined as follows.
\begin{equation}
\nonumber
    L_d(M,X,X^\prime)=\sup_{S\subseteq{\mathcal{S}}}\left|\log\frac{\Pr{(M(X)\in S))}}{\Pr{(M(X^\prime)\in S}))}\right|
\end{equation}
Then, mechanism $M$ satisfies $\epsilon$-differential privacy iff $L_d(M,X,X^\prime)\leq{\epsilon}$ for any neighboring datasets $X,X^\prime$.
\end{definition}

$\epsilon$-differential privacy guarantees that the outputs of mechanism $M$ are similar (i.e., privacy loss is bounded up to $\epsilon$) when the inputs are neighboring. In other words, from the output of algorithm $M$, it is difficult to infer what a single record is due to the definition of the neighboring datasets. In this study, we apply differential privacy to a setting of a location on a road network.

\subsection{Geo-indistinguishability}
\label{subsec:geo-i}
Here, we describe the definition of geo-indistinguishability (Geo-I)~\cite{geo-i}. Let $\mathcal{X}$ be a set of locations. Intuitively, a mechanism $M$ that achieves Geo-I guarantees that $M(x)$ and $M(x^\prime)$ are similar to a certain degree for any two locations $x,x^\prime\in\mathcal{X}$. This means that even if an adversary obtains an output from this mechanism, a true location will be indistinguishable from other locations to a certain degree. When $\mathcal{X}\subseteq{\mathbb{R}^2}$, $\epsilon$-Geo-I is defined as follows~\cite{geo-i}.
\begin{definition}[$\epsilon$-geo-indistinguishability~\cite{geo-i}]
Let $\mathcal{Z}$ be a set of query outputs.
A mechanism $M:\mathcal{X}\to\mathcal{Z}$ satisfies $\epsilon$-Geo-I iff $\forall{x,x^\prime}\in\mathcal{X}$:
\begin{equation}
\nonumber
    \sup_{S\in{\mathcal{Z}}}\left|\log\frac{\Pr(M(x)\in S)}{\Pr(M(x^\prime)\in S)}\right|\leq\epsilon d_e(x,x^\prime)
\end{equation}
where $d_e$ is the Euclidean distance.
\end{definition}

\subsubsection{Mechanism satisfying $\epsilon$-Geo-I}
\label{subsubsec:plm}
The authors of~\cite{geo-i} introduced a mechanism called the planar Laplace mechanism (PLM) to achieve $\epsilon$-Geo-I. The probability distribution generated by PLM is called the planar Laplace distribution and---as its name suggests---is derived from a two-dimensional version of the Laplace distribution as follows:
\begin{equation}
\nonumber
    \Pr(PLM_\epsilon(x)=z) = \frac{\epsilon^2}{2\pi}\mathrm{e}^{-\epsilon d_e(x,z)}
\end{equation}
where $x,z\in\mathcal{X}$.

\subsection{Problem Statement}
\label{subsec:problem}
We consider a perturbation mechanism to improve the tradeoff between utility and privacy by taking advantage of road networks. We assume that the LBSs work on road networks (e.g., UBER), that users are located on road networks, and that LBS providers expect to receive a location on a road network.

We model a road network as an undirected weighted graph $G=(V,E)$ and locations on the road network as the vertices $V$ that are on the Euclidean plane $\mathbb{R}^2$. Each edge in $E$ represents a road segment and the weight of the edge is the length of the road segment.
Then, the distance is the shortest path length $d_s$ between two nodes.
Here, the following inequality holds for any two vertices on $v,v^\prime\in\mathcal{V}$.
\begin{equation}
\label{equ:shortest-path}
d_e(v,v^\prime)\leq d_s(v,v^\prime)
\end{equation}
where $d_e$ is the Euclidean distance.

We assume that a user is located at a location on a road network $v\in V$, sends the location once to receive service from an untrusted LBS, and that an adversary knows that the user is on the road network. The user needs to protect his/her privacy on his/her own device using a perturbation mechanism $M:V\to\mathcal{W}$ where $\mathcal{W}\subseteq V$. This is the same setting as the setting of the local differential privacy~\cite{kasiviswanathan2011can}.

%To evaluate privacy, we use two kinds of notions: empirical privacy (Equation~\ref{equ:privacygain}) and indistinguishability (i.e., our proposed notion). Empirical privacy measures the level of privacy protection against a certain adversary. In this paper, empirical privacy means \privacygain\ and the indistinguishability is a global notion of location privacy on road networks, $\epsilon$-\roadi, which is our proposal.
%To evaluate utility, We use \utilityloss\ (Equation~\ref{SQL_def}).

%%% takagi 12/19
Goals of this paper are to formally define privacy of locations on road networks and to achieve a better tradeoff between privacy and utility by considering road networks than existing method~\cite{geo-i} based on the Euclidean plane.
%%% end

The main notations used in this paper are summarized in Table \ref{notations}.
\begin{table*}[t]
 \centering
  \begin{tabular}{cl}
   \hline
   Symbol & \multicolumn{1}{c}{Meaning} \\
   \hline \hline
    $u,a$ & A user and an adversary.\\
    $\mathbb{R}$ & Set of real numbers.\\
    $\mathcal{Z}$ & Set of outputs.\\
    $G=(V,E)$ & Weighted undirected graph that represents a road network.\\
    $V$ & Set of vertices.\\
    $E$ & Set of edges. A weight is the distance on the road segment connecting two vertices.\\
    $\mathcal{W}\subseteq V$ & Set of vertices of outputs.\\
    $v,v^\prime,\hat{v}$ & On a road network, a true vertex, a perturbed vertex and an inferred vertex.\\
    $x,x^\prime,\hat{x}$ & On the Euclidean plane, a true location, a perturbed location and an inferred location.\\
    $\pi_u(x)$ & The probability that user $u$ is at location $x$.\\
    $\pi_a(x)$ & Adversary $a$'s knowledge about user's location that represents the probability of being at location $x$.\\
    $M$ & A mechanism. Given a location, $M$ outputs a perturbed location.\\
    $d_e(x,x^\prime)$ & An Euclidean distance between $x$ and $x^\prime$.\\
    $d_s(v,v^\prime)$ & The shortest distance between $v$ and $v^\prime$ on a road network.\\
    $h$ & Inference function that represents inference of an adversary.\\
    $f$ & Post-processing function.\\
   \hline
   \vspace{3pt}
  \end{tabular}
 \caption{Summary of notation.}
 \label{notations}
\end{table*}

\section{Geo-graph-indistinguishability}

In this section, we propose a new definition of location privacy on road networks, called Geo-Graph-Indistinguishability (GG-I). We first formally define GG-I. Then, we clarify the relationship between Geo-I and GG-I. In the following subsections, we describe the reason why GG-I restricts the output range and characteristics that GG-I inherits from $d_\mathcal{X}$-privacy~\cite{broad_dp}.

\subsection{Definition}
We assume that a graph $G=(V,E)$ representing a road network is given.
First, we introduce the privacy loss of a location on a road network as follows.
\begin{definition}[privacy loss of a location on a road network]
Given a mechanism $M:V\to\mathcal{Z}$ and $v,v^\prime\in V$, privacy loss of a location on a road network is as follows:
\begin{equation}
\nonumber
L(M,v,v^\prime)=\sup_{S\subseteq\mathcal{Z}}\left|\log\frac{\Pr(M(v)\subseteq S)}{\Pr(M(v^\prime)\subseteq S)}\right|
\end{equation}
\end{definition}
Intuitively, privacy loss measures how much different two outputs are for two inputs $v$ and $v^\prime$. If the privacy loss value $L(M,v,v^\prime)$ is small, an adversary who sees an output $M(v)$ cannot distinguish the true location from $v$ and $v^\prime$, which is the basic notion of differential privacy described in~\sect~\ref{subsec:differentialprivacy}. In the same way that differential privacy guarantees the indistinguishability of a record in a database, our notion guarantees the indistinguishability of a location. 
Given $\epsilon \in \mathbb{R^+}$, we define $\epsilon$-geo-graph-indistinguishability as follows.
\begin{definition}{($\epsilon$-geo-graph-indistinguishability)}
Mechanism $M:V\to V$ satisfies $\epsilon$-GG-I iff $\forall v,v^\prime\in V$, 
\begin{equation}
\nonumber
    L(M,v,v^\prime)\leq \epsilon d_s(v,v^\prime)
\end{equation}
where $d_s$ is the shortest path length between two vertices.
\end{definition}
Intuitively, $\epsilon$-GG-I constrains any two outputs of a mechanism to be similar when the two inputs are similar, that is, they will represent close vertices. In other words, two distributions of two outputs are guaranteed to be similar. The degree of similarity of two probability distributions is $\epsilon d_s(v,v^\prime)$. From this property, an adversary who obtains an output of the mechanism cannot distinguish the true input $v$ from other vertices $v^\prime$ according to the value of $\epsilon d_s(v,v^\prime)$. In particular, a vertex close to the true vertex cannot be distinguished. Moreover, $\epsilon$-GG-I constrains the output range to the vertices of the graph because an output consisting of locations other than those on the road network may cause empirical privacy leaks. This constraint prevents such kind of privacy leak. We provide additional explanation of this concept in \sect~\ref{subsec:range}.
The definition can be also formulated as follows:
\begin{equation}
\nonumber
    \forall v,v^\prime\in V, \forall W\subseteq V,\frac{\Pr(M(v)\subseteq W)}{\Pr(M(v^\prime)\subseteq W)} \leq \mathrm{e}^{\epsilon d_s(v,v^\prime)}
\end{equation}
This formulation implies that GG-I is an instance of $d_\mathcal{X}$-privacy~\cite{broad_dp} proposed by Chatzikokolakis et al. as are Geo-I and differential privacy. Chatzikokolakis et al. showed that an instance of $d_\mathcal{X}$-privacy guaranteed strong privacy property as shown in \sect~\ref{subsec:chara}.

\subsection{Relationship between Geo-I and GG-I}
\label{subsec:analyze_relationship}
Geo-I~\cite{geo-i} defines location privacy on the Euclidean plane (see \sect~\ref{subsec:geo-i} for details). Here, we explain the relationship between Geo-I and GG-I. To show the relationship, we introduce the following lemma.
\begin{lemma}[Post-processing theorem of Geo-I.]
\label{lemma:postprocess}
If a mechanism $M:\mathcal{X}\to\mathcal{Z}$ satisfies $\epsilon$-Geo-I, a post-processed mechanism $f\circ M$ also satisfies $\epsilon$-Geo-I for any function $f:\mathcal{Z}\to\mathcal{Z}^\prime$.
\end{lemma}
We refer readers to the appendix for the proof. Intuitively, this means that Geo-I does not degrade even if the output is mapped by any function. Moreover, if a mechanism $M:\mathcal{X}\to\mathcal{Z}$ satisfies $\epsilon$-Geo-I, the following inequality holds for any two vertices $v,v^\prime\in V$ from Inequality~(\ref{equ:shortest-path}).
\begin{equation}
\nonumber
\begin{split}
    \sup_{S\in{\mathcal{Z}}}\left|\log\frac{\Pr(M(v)\in S)}{\Pr(M(v^\prime)\in S)}\right|\leq &\epsilon d_e(v,v^\prime)\\\leq
    &\epsilon d_s(v,v^\prime)
\end{split}
\end{equation}
From this inequality and Lemma~\ref{lemma:postprocess}, we can derive the following theorem.
\begin{theorem}
If a mechanism $M$ satisfies $\epsilon$-Geo-I, $f\circ M$ satisfies $\epsilon$-GG-I, where $f:\mathcal{Z}\to V$ is any mapping function to a vertex of the graph.
\end{theorem}
This means that a mechanism that satisfies $\epsilon$-Geo-I can always be converted into a mechanism that satisfies $\epsilon$-GG-I by post-processing.
We note that the reverse is not always true. That is, GG-I is a relaxed version of Geo-I through the use of the metric $d_s$, allowing for us to create a mechanism that outputs a useful location. We refer to \sect~\ref{subsec:privacy_geoi_geogi} for details.

For example, the planar Laplace mechanism (\plm)~(\sect~\ref{subsubsec:plm}) satisfies $\epsilon$-Geo-I. Because Outputs of PLM consist locations other than locations on a road network, it may cause empirical privacy leaks as described in the next section; this is because PLM does not satisfy $\epsilon$-GG-I. 
$f\circ \plm$ satisfies $\epsilon$-GG-I and prevents this privacy leaks if $f$ is a mapping function to a vertex of a graph. For utility, we can use a mapping function that maps to the nearest vertex; we call this mechanism the Planar Laplace Mechanism on a Graph (\plmg).

\subsection{Output Range from a Privacy Perspective}
\label{subsec:range}
There are two reasons why $\epsilon$-GG-I restricts output range to vertices of the graph.
First, LBSs that operate over road networks expect to receive a location on a road network as described in \sect~\ref{subsec:problem}.

Second, because road networks are public information, outputting a location outside the road network may cause empirical privacy leaks. We empirically show that an adversary who knows the road network can perform a more accurate attack than can one who does not know the road network; a post-processed mechanism protects privacy from this type of attack.
To show this, we evaluate the empirical privacy gain AE of two kinds of mechanisms PLM and PLMG against the two kinds of adversaries. 
%We use function $f$, which maps to the nearest location on road networks as post-processing of \plmg.

\begin{figure}[t]
 \begin{minipage}{0.4\hsize}
   \centering\includegraphics[width=\hsize]{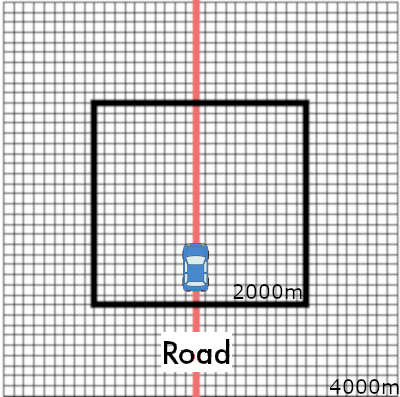}
  \caption{A synthetic map. The red line represents a road, and a user is located inside the black frame.}
  \label{fig:syn_map}
 \end{minipage}
\hfill
 \begin{minipage}{0.6\hsize}
   \centering\includegraphics[width=\hsize]{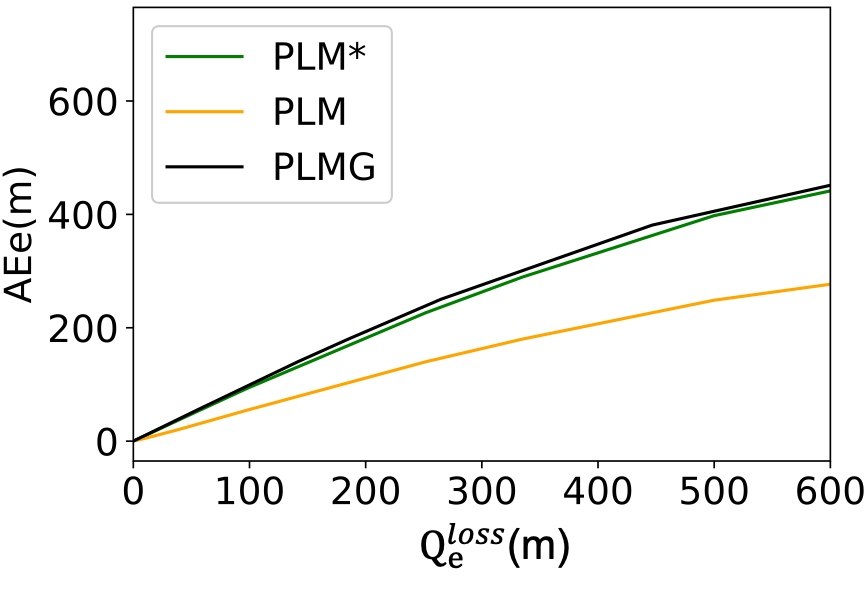}
  \caption{AE of each mechanism with respect to Q$^{loss}$ with the Euclidean distance, that is AE$_e$ and Q$^{loss}_e$. PLM$^*$ represents the AE of PLMG against an adversary who does not know the road network.}
  \label{fig:output_outside}
 \end{minipage}
\end{figure}

For simplicity, we use a simple synthetic map illustrated in \fig~\ref{fig:syn_map}. This map consists of 1,600 squares each of which has a side length of \SI{100}{m}; that is, the area dimensions are \SI{4000}{m} * \SI{4000}{m}, and each lattice point has a coordinate. The centerline represents a road where a user is able to be located, and the other areas represent locations where a user must not be, such as the sea. 
In this map, we evaluate the empirical privacy gain AE of the two mechanisms against two kinds of adversaries with the same utility loss Q$^{loss}$. We use Euclidean distance as the metric of AE and Q$^{loss}$, denoted by AE$_e$ and Q$^{loss}_e$, respectively. 

\fig~\ref{fig:output_outside} shows the results. 
PLM represents the empirical privacy gain AE against an adversary who knows the road network, while PLM$^*$ represents the AE against an adversary who dose not know the road network. Comparing PLM with PLM$^*$, the adversary can more accurately infer the true location by considering the road network.
The AE of PLMG is higher than the AE of PLM and almost identical to the AE of PLM$^*$. 
By restricting the output to locations on the road network, the adversary cannot improve the inference of the true location because no additional information exists. In other words, post-processing to a location on road networks strengthens the empirical privacy level against an adversary who knows the road network.

\subsection{Characteristics}
\label{subsec:chara}
\rni\ is an instance of $d_\mathcal{X}$-privacy~\cite{broad_dp}, which is a generalization of differential privacy with the following two characteristics that show strong privacy protection.

\subsubsection{Hiding function}
The first characteristic uses the concept of a hiding function $\phi:V\to V$, which hide a secret location by mapping to the other location.
For any hiding function and a secret location $v\in V$, when an attacker who has a prior distribution that includes information about the user's location obtains each output $o = M(v)$ and $o^\prime = M(\phi(v))$ of a mechanism that satisfies $\epsilon$-GG-I, the following inequality holds for each posterior distribution:
\begin{equation}
\nonumber
    \left|\log\frac{\Pr(v|o)}{\Pr(v|o^\prime)}\right|\leq 2\epsilon \sup_{v\in V}d_s(v,\phi(v))
\end{equation}
This inequality guarantees that the adversary's conclusions are the same (up to $2\epsilon \sup_{v\in V}d_s(v,\phi(v))$) regardless of whether $\phi$ has been applied to the secret location.
\subsubsection{Informed attacker}
\label{subsubsec:informed}
The other characteristic can be shown by the ratio of a prior distribution and posterior distribution, which is derived by obtaining an output of the mechanism. By measuring this value, we can determine how much the adversary has learned about the secret. We assume that an adversary (informed attacker) knows that the secret location is in $N\subseteq{V}$. When the adversary obtains an output of the mechanism, the following inequality holds for the ratio of his prior distribution $\pi_{|N}(v)=\pi(v|N)$ and its posterior distribution $p_{|N}(v|o)=p(v|o,N)$:
\begin{equation}
\nonumber
    \log\frac{\pi_{|N}(v)}{p_{|N}(v|o)}\leq \epsilon \max_{v,v^\prime\in N}d_s(v,v^\prime)
\end{equation}
Intuitively, this means that the more the adversary knows about the actual location, the less he will be able to learn about the location from an output of the mechanism.\par

\section{A Mechanism to Achieve Geo-Graph-Indistinguishability}
Here, we assume that a graph $G=(V,E)$, which represents a road network, is given, and we propose a mechanism that satisfies GG-I, which we call the Graph-Exponential Mechanism (GEM). 
Second, we explain the implementation of GEM.
Third, we describe an advantage and an issue of \gem\ caused by not satisfying Geo-I.

\subsection{Graph-Exponential Mechanism}
PLMG (\sect~\ref{subsec:analyze_relationship}) satisfies GG-I, but PLMG does not take advantage of the structures of road networks to output useful locations. Here, we propose a mechanism that considers the structure of road networks so that the mechanism can output more useful locations. Given the parameter $\epsilon\in\mathbb{R^+}$ and a set of outputs $\mathcal{W}\subseteq{V}$, $GEM_\epsilon$ is defined as follows.
\begin{definition} 
$GEM_\epsilon$ takes $v\in V$ as an input and outputs $o\in{\mathcal{W}}$ with the following probability.
\begin{equation}
\label{equ:gem}
    \Pr(GEM_\epsilon(v)=o) = \alpha(v)\mathrm{e}^{-\frac{\epsilon}{2}d_s(v,o)}
\end{equation}
where $\alpha$ is a normalization factor $\alpha(v)=(\sum_{o\in \mathcal{W}}\mathrm{e}^{-\frac{\epsilon}{2} d_s(v,o)})^{-1}$.
\end{definition}
This mechanism employs the idea of an exponential mechanism~\cite{mcsherry2007mechanism} that is one of the general mechanisms for differential privacy.
Because this mechanism capitalizes on the road network structure by using the metric $d_s$, it can achieve higher utility for LBSs over road networks than can PLMG as shown in \sect~\ref{sec:experiments}.
\begin{theorem}
GEM$_\epsilon$ satisfies $\epsilon$-\rni.
\end{theorem}
We refer readers to the appendix for the proof.

\subsection{Computational complexity of \gem}
Since we assume that LBS providers are untrusted and there is no trusted server, a user needs to create the distribution and sample the perturbed location according to the distribution locally. Here, we explore a method to accomplish this and the issues that can be caused by the number of vertices.

GEM consists of three phases: (i) obtain the shortest path lengths to all vertices from the user's location. (ii) compute the distribution according to Equation~(\ref{equ:gem}). (iii) sample a point from the distribution. We show the pseudocode of GEM in~Algorithm~\ref{alg:gem}.

\begin{algorithm}                      
\caption{Graph-exponential mechanism.}         
\label{alg:gem}                          

\begin{algorithmic}      

\newlength\myindent
\setlength\myindent{2em}
\newcommand\bindent{%
    \begingroup
    \setlength{\itemindent}{\myindent}
    \addtolength{\algorithmicindent}{\myindent}
}
\newcommand\eindent{\endgroup}

\REQUIRE {Privacy parameter $\epsilon$, true location $v$, graph $G=(V,E)$, output range $\mathcal{W}\subseteq{V}$.}
\ENSURE {Perturbed location $w$.}
\STATE \textbf{(i)} $d_s(v,\cdot) \Leftarrow Dijkstra(G=(V,E), v)$ 
\STATE \textbf{(ii)} Compute the distribution:
\bindent
\FOR{$v$ in $\mathcal{W}$}
\STATE  $\Pr(GEM(v)=w) \Leftarrow \alpha(v)\mathrm{e}^{-\epsilon d_s(v,w)/2}$
\ENDFOR
\eindent
\STATE \textbf{(iii)} $w \sim \Pr(GEM(v)=w)$
\STATE return $w$
\end{algorithmic}
\end{algorithm}

We next analyze the computational complexity of each phase.
For phase (i), \gem\ computes the shortest path lengths to the other nodes from $v$. The computational complexity of this operation is $O(|E|+|V|\log |V|)$ by using Fibonacci heap, where $|V|$ is the number of nodes and $|E|$ is the number of edges.
This level of computational complexity does not cause a problem, but on road networks, a fast algorithm computing the shortest path length has been studied for large numbers of graph vertices; we refer the reader to~\cite{akiba2014fast} that may be applied to our algorithm. Phase (ii) has no computational problem because its computational complexity is $O(|V|)$. In phase (iii), when the number of vertices is much larger than we expect, we may not be able to effectively sample the vertices according to the distribution. This problem has also been studied and is known as consistent weighted sampling (CWS); we refer the reader to~\cite{consistent-weighted-sampling,wu_improved_2017}. We believe that these studies can be applied to our algorithm and can be computed even when the number of vertices is somewhat large.

\subsection{Privacy with Respect to Euclidean distance}
\label{subsec:privacy_geoi_geogi}
As described in \sect~\ref{subsec:analyze_relationship}, PLMG satisfies $\epsilon$-Geo-I and $\epsilon$-GG-I, but GEM satisfies only $\epsilon$-GG-I.
This is because GG-I is a relaxed definition of Geo-I that allows a mechanism to output a more useful perturbed location. Therefore, GEM shows better utility as shown in experiments of \sect~\ref{sec:experiments}.
It is worth investigating whether this relaxation weakens the privacy protection guarantees. In short, GG-I has no privacy protection guarantees with respect to Euclidean distance; thus, if a user is using a mechanism that satisfies GG-I to location privacy, the adversary may easily be able to distinguish the user's location from other locations even when those other locations are close to the user's location based on Euclidean distance. In what follows, we demonstrate this fact using the notion of true probability (TP). The probability that an adversary can distinguish a user's location is
\begin{equation}
\begin{split}
\nonumber
    &\tp(\pi_u,M,h) \\ &= \sum_{v,\hat{v}\in{\mathcal{V}},o\in{\mathcal{W}}}\pi_u(v)\Pr(M(v)=o)\Pr(h(o)=v^\prime)\delta(v,\hat{v})
\end{split}
\end{equation}
where $\delta(\hat{v},v)$ is a function that returns $1$ if $\hat{v}=v$ holds; otherwise, it returns $0$. TP is the expected probability with which an adversary can remap a perturbed location to the true location. 

\begin{figure}[t]
 \begin{minipage}{0.5\hsize}
   \centering\includegraphics[width=\hsize]{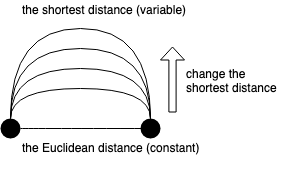}
  \caption{Each graph has a different shortest path length with the same Euclidean distance.}
  \label{fig:graph_tp}
 \end{minipage}
\hfill
 \begin{minipage}{0.5\hsize}
   \centering\includegraphics[width=\hsize]{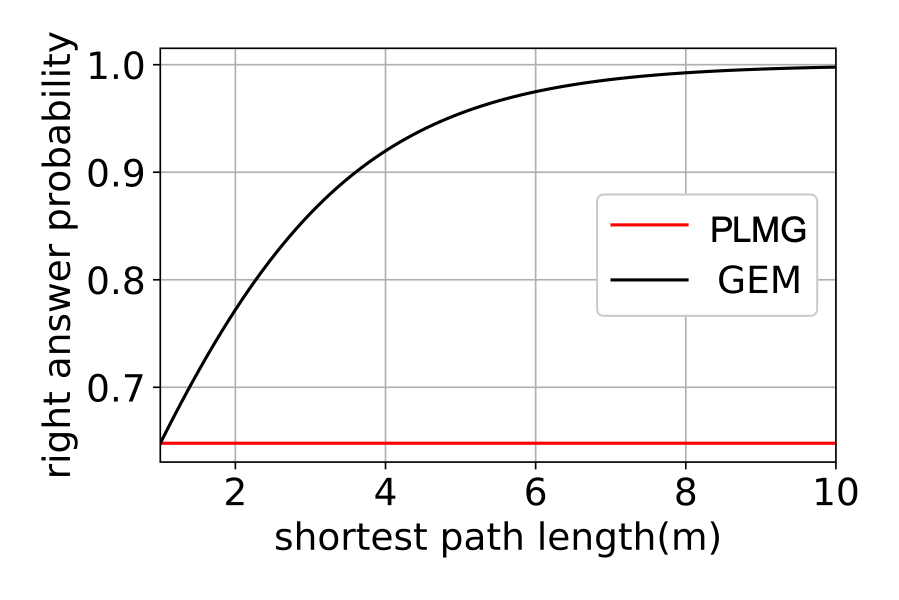}
  \caption{\tp\ according to \gem\ and \plmg.}
  \label{fig:ggi_weak_img}
 \end{minipage}
\end{figure}

We assume a set of graphs, each of which has only two vertices. The Euclidean distances between the vertices are the same for all the graphs, but weights of the edges between them are different for each graph (\fig~\ref{fig:graph_tp}). Next, we assume that each prior of a user's location is a uniform distribution on two vertices of this graph, and we compute TP of PLMG and GEM. \fig~\ref{fig:ggi_weak_img} shows the change in TP when the weight (that is, the shortest path length) changes. Due to the guarantee of the Euclidean distance of Geo-I, PLM does not degrade TP even when the shortest path length changes, however, since GG-I does not have a guarantee of the Euclidean distance, GEM significantly degrades TP, which means that the adversary can discover the user's true location.\par

A mechanism satisfying $\epsilon$-GG-I can achieve better utility than can a mechanism satisfying Geo-I by guaranteeing privacy protection in terms of the shortest distance on road networks instead of the Euclidean distance. This idea comes from the interpretation of privacy; in this paper, we assume that privacy can be interpreted as the shortest distance on road networks. Therefore, GG-I may not be suitable for protecting location privacy when the privacy needs to be interpreted as Euclidean distance, e.g., weather conditions, where a wide range of locations need to be protected.

\begin{figure}[t]
 \begin{minipage}{0.5\hsize}
   \centering\includegraphics[width=\hsize]{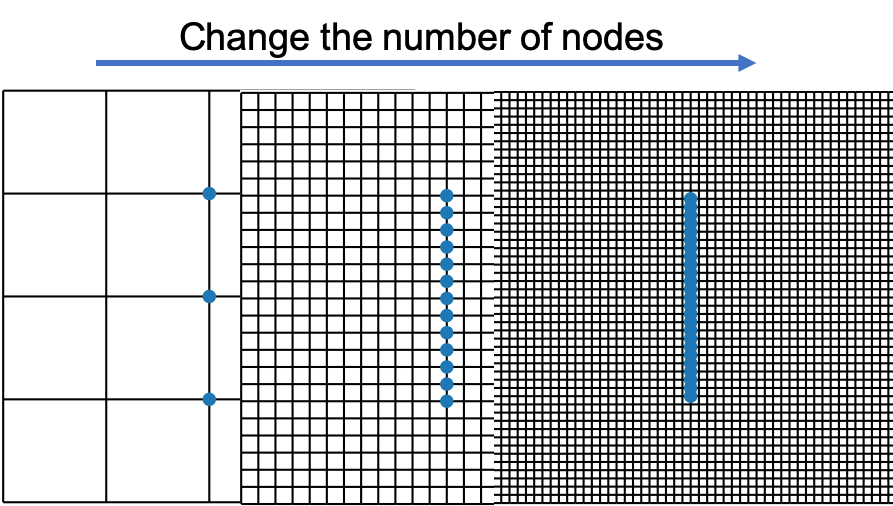}
  \caption{Points represent graphs nodes, which we use as the input and output of mechanisms. There are edges between neighboring nodes. The side length of each square is \SI{1000}{m}.}
  \label{fig:square_graphs}
 \end{minipage}
\hfill
 \begin{minipage}{0.5\hsize}
   \centering\includegraphics[width=\hsize]{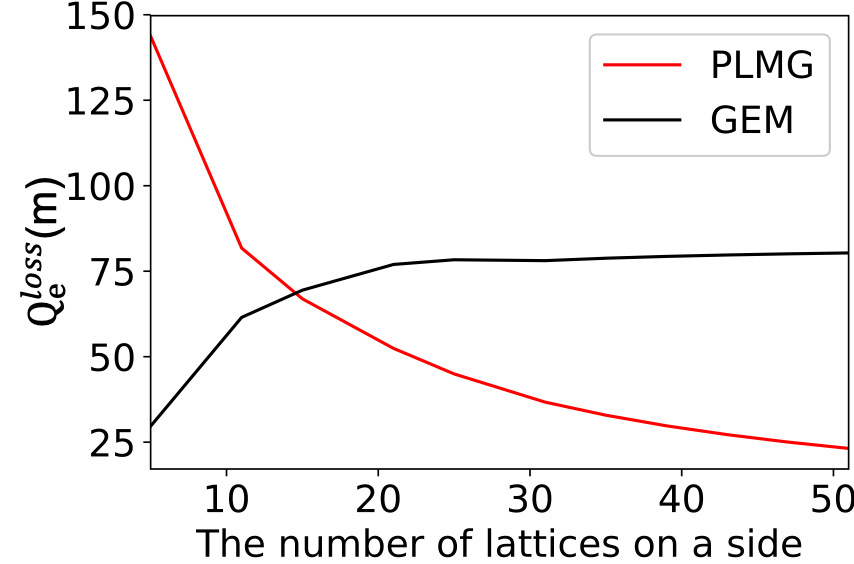}
  \caption{Utility loss when changing the number of nodes with $\epsilon=0.01$.}
  \label{fig:lattice_sql}
 \end{minipage}
\end{figure}

\subsection{Utility Comparison with PLMG}
Both GEM$_\epsilon$ and PLMG$_\epsilon$ satisfy $\epsilon$-GG-I, which means that both guarantee the same indistinguishability. However, outputs of GEM and PLMG are created from different distributions: the continuous distribution with post-processing and the discrete distribution, respectively. Here, we explore the change in utility yielded by their difference; consequently, we use synthetic graphs~(Blue points in \fig~\ref{fig:square_graphs}) whose shortest path lengths and Euclidean distances between two nodes are identical to exclude the difference caused by the variations in the adopted metrics---that is, graphs that have the shape of a straight line on a Euclidean plane. We prepare several graphs by changing the number of nodes while fixing the length of the entire graph. \fig~\ref{fig:lattice_sql} shows the utility loss (i.e., Q$^{loss}$) of GEM and PLMG with $\epsilon=0.01$ for each graph. As shown, the Q$^{loss}$ of GEM increases as the number of nodes increases, while the Q$^{loss}$ of PLMG decreases. This is also the result with other $\epsilon$ values. PLMG is post-processed by mapping to the nearest node, so when few nodes exists near the output of PLM, PLMG cannot output a useful location because the mapping to the location may be distant from the input. Conversely, GEM cannot efficiently output a useful location when there are many nodes because GEM needs to distribute the probabilities to distant nodes. As mentioned in \sect~\ref{subsec:problem}, road networks are generally discretized by graphs, and it can be said that GEM is an appropriate mechanism on road networks. We will also show the effectiveness of GEM compared with PLMG according to the utility in the real-world road networks. We refer to \sect~\ref{subsec:exp_gem_plmg} for details.

\section{Analyzing the Performance of GEM and Optimizing Range}
GEM requires output $z$ to be on a road network but require nothing else for the output range. This means that an optimal output range exists for privacy and utility.
In this section, first we apply Q$^{loss}$ and AE to a location setting on road networks. Then, we propose the performance criteria (PC) which represents the tradeoff between the privacy and  the utility.
Next, we formalize an optimization problem for the PC.
Finally, we propose a greedy algorithm to solve the optimization problem in an acceptable amount of time.

\subsection{Performance of a Mechanism on a Road Network}
While the $\epsilon$ of GG-I indicates the degree of indistinguishability between a real and perturbed location, it does not indicate the performance of a mechanism w.r.t its utility for some user and empirical privacy against some adversary. 
Therefore, we introduce the two notions Q$^{loss}_s$ and AE$_{s}$ by applying Q$^{loss}$ and AE (\sect~\ref{subsubsec:aeandsql}) to the setting of road networks. We provide their definitions below.

\begin{equation}
\nonumber
    Q^{loss}_s(\pi_u,M) = Q^{loss}(\pi_u,M, d_s) 
\end{equation}
\begin{equation}
\nonumber
    AE_{s}(\pi_a,M,h) = AE(\pi_a, M, h, d_s)
\end{equation}

Intuitively, Q$^{loss}_s$ is the expected distance on road networks between the true locations and perturbed locations, while AE$_s$ is the expected distance on road networks between the true locations and the locations inferred by an adversary. 
In the following, we let Q$^{loss}$ and AE denote Q$^{loss}_s$ and AE$_s$, respectively.
We note that, as opposed to $\epsilon$, AE changes according to the assumed adversary (i.e., the specific attack method and prior distribution). 
However, because AE increases as Q$^{loss}$ increases (e.g., a mechanism that outputs a distant location will result in high AE but also high Q$^{loss}$), using only AE as a performance criterion for a mechanism is not appropriate.
Then, we define a new criterion to measure the performance of a mechanism against an assumed adversary, which we call the performance criterion (PC).

\begin{equation}
\nonumber
    PC=AE/Q^{loss}
\end{equation}
Intuitively, against an assumed adversary, PC represents the size of AE with respect to the Q$^{loss}$. In other words, PC measures the utility/privacy tradeoff.
For example, if an adversary with an optimal attack~\cite{shokri-strategy} cannot infer the true location at all (i.e., the adversary infers the pseudolocation as the true location), the mechanism can be considered as having the highest performance ($PC=1$). Conversely, the mechanism performs worst ($PC=0$) if the adversary can always infer the true location.

\subsection{Objective Functions}
Here, we propose an objective function to find the optimal output range of GEM with respect to the performance. We assume that the prior distribution of a user is given and adversary knows the prior distribution. An example of this is shown in \sect~\ref{subsubsec:scenario}. If the prior distribution is not give, we can use uniform distribution for the general user.

Then, we can compute AE and Q$^{loss}$ by assuming an inference function (we refer to \sect~\ref{subsubsec:userandadv} for detail). 
We use a posterior distribution given the pseudolocation $o$ as the inference function $h$. 
Then, given an output range $\mathcal{W}$, the PC of GEM with the output range $\mathcal{W}$ is formulated as follows:

\begin{equation}
\nonumber
    \frac{\sum_{v,\hat{v}\in V,o\in\mathcal{W}}\pi_u(v)\Pr(\gem_\mathcal{W}(v)=o)p(\hat{v}|o)d_s(v,\hat{v})}{\sum_{v\in V,o\in\mathcal{W}}\pi_u(v)\Pr(\gem_\mathcal{W}(v)=o)d_s(v,o)}
\end{equation}
where GEM$_\mathcal{W}$ denotes GEM with the output range $\mathcal{W}$.
Then, the objective function against the adversary can be formulated as follows.
\begin{maxi}|l|
{\mathcal{W}\subseteq{V}}{\mc_{\mathcal{W}}}{}{}
\nonumber
\end{maxi}
where PC$_\mathcal{W}$ is the PC of GEM$_\mathcal{W}$.
Here, GEM with the optimized output range is considered to show the best tradeoff against the adversary, but it can fail to be useful (i.e. large Q$^{loss}$) because Q$^{loss}$ has no constraints; consequently we add the following constraint to Q$^{loss}$.

\begin{maxi}|l|
{\mathcal{W}\subseteq{V}}{\mc_\mathcal{W}}{}{}
\addConstraint{Q^{loss}_{\mathcal{W}} \leq \theta}
\nonumber
\end{maxi}
where Q$^{loss}_\mathcal{W}$ is the Q$^{loss}$ of GEM$_\mathcal{W}$.
The optimal GEM shows the best tradeoff in GEM with an output range that shows a better Q$^{loss}$ than $\theta$. We set Q$^{loss}_{\mathcal{W}_0}$ to $\theta$ so that the utility does not degrade by the optimization.

\subsection{Algorithm to Find an Approximate Solution}
\label{subsec:tactic}
Because the number of combinations for the output range is $2^{|V|}$, we cannot compute all combinations to find the optimal solution for the optimized problem in an acceptable amount of time; therefore, we propose a greedy algorithm that instead finds approximate solutions. The pseudocode for this algorithm is listed in Algorithm~\ref{alg:opt}. The constraint function is a function that returns a value indicating whether the constraint holds or does not hold.

\begin{algorithm}             
\caption{Finding a local solution.}         
\label{alg:opt}                          
\begin{algorithmic}[1]
\REQUIRE {Privacy parameter $\epsilon$, graph $G=(V,E)$ objective function $f$, constraint function $c$, initial output range $\mathcal{W}_0$.}
\ENSURE {Output range $\mathcal{W}$.}
\WHILE{True}
\STATE $obj \Leftarrow f(GEM_{\mathcal{W}_o})$

\FOR{$v$ in $V$}
\STATE $\mathcal{W}^\prime \Leftarrow \mathcal{W}\setminus \{v\}$
\STATE $obj^\prime \Leftarrow f(GEM_{\mathcal{W}^\prime})$
\STATE $cons \Leftarrow c(GEM_{\mathcal{W}^\prime})$
\IF{$obj^\prime - obj < 0$ and cons}
\STATE $\mathcal{W} \Leftarrow W^\prime$
\STATE $obj \Leftarrow obj^\prime$
\ENDIF
\ENDFOR
\IF{$\mathcal{W}_0 = \mathcal{W}$}
\STATE break
\ENDIF
\ENDWHILE
\STATE return $\mathcal{W}$
\end{algorithmic}
\end{algorithm}

First, we start with a given initial output range $\mathcal{W}_0$. Next, we compute a value of the objective function of the output range with one node removed. We remove that node if the objective function improves and the constraint holds. We repeat this procedure until the objective function converges, which has a computational complexity of $O(|\mathcal{W}_0|^2)$ in the worst case when the computational complexity of the objective function is $O(1)$.
As a rule of thumb, the main loop (line 2 of Algorithm \ref{alg:opt}) likely completes in only a small number of iterations.
However, the computational complexity of PC is $O(|V|^2|\mathcal{W}_0|)$, so the overall computational $O(|V|^2|\mathcal{W}_0|^3)$.
Therefore, when $|\mathcal{W}_0|$ is large, this computational complexity is not acceptable.
In the following, we propose a way of providing $\mathcal{W}_0$.
\subsubsection{Initialization of $\mathcal{W}$}
PC increases when Q$^{loss}$ decreases, so we propose to first optimize output range according to Q$^{loss}$, which is computed in the small computational complexity.
The optimization problem is as follows:
\begin{mini}|l|
{\mathcal{W}\subseteq{V}}{Q^{loss}_{\mathcal{W}}}{}{}
\nonumber
\end{mini}
Q$^{loss}_{\mathcal{W}\setminus{v}}$ can be computed using Q$^{loss}_{\mathcal{W}}$ in the computational complexity of $O(|V|)$.
Therefore, we can obtain an approximate solution according to this optimization problem using Algorithm~\ref{alg:opt} with the initial output range $V$ in the computational complexity of $O(|V|^3)$ in the worst case. As described above, the main loop likely completes in only a small number of iterations, so we can complete this algorithm in the computational complexity of $O(|V|^2)$ in the most case, and this is acceptable even when $|V|$ is somewhat large.
We use this output range as the initial output range of Algorithm~\ref{alg:opt}.

\subsection{Optimization Examples}

\begin{figure}[t]
 \begin{minipage}{0.49\hsize}
   \centering\includegraphics[width=\hsize]{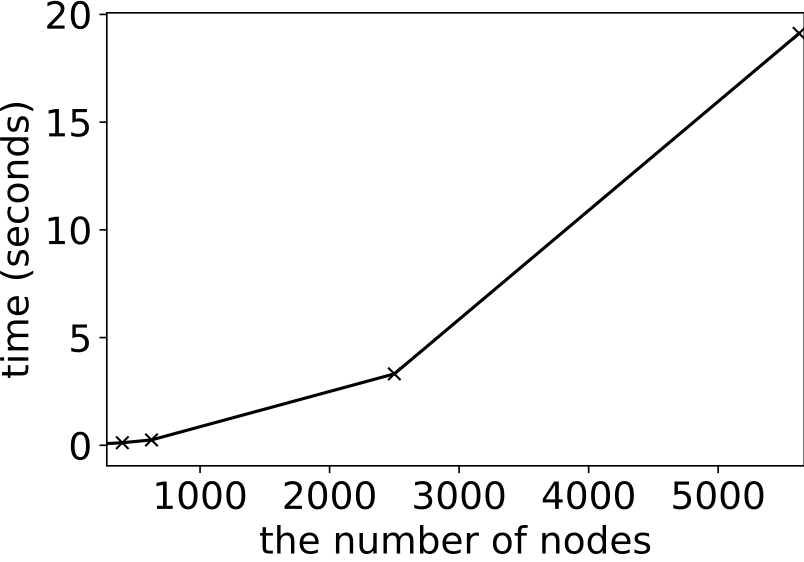}
  \caption{The relationship between the number of nodes and time required for the optimization.}
  \label{fig:time}
 \end{minipage}
 \begin{minipage}{0.49\hsize}
   \centering\includegraphics[width=\hsize]{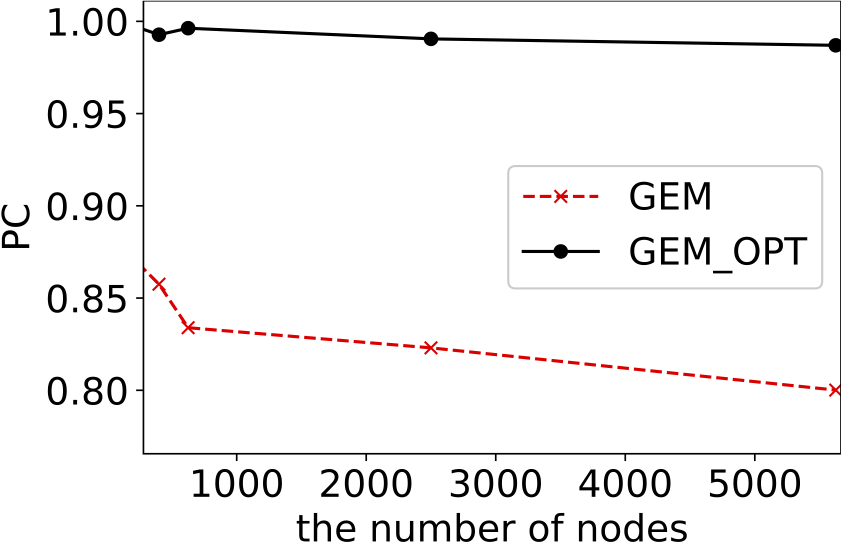}
  \caption{The relationship between PC and the number of nodes.}
  \label{fig:PC_nodes}
 \end{minipage}
\end{figure}

\begin{figure}[t]
 \begin{minipage}{0.5\hsize}
   \centering\includegraphics[width=\hsize]{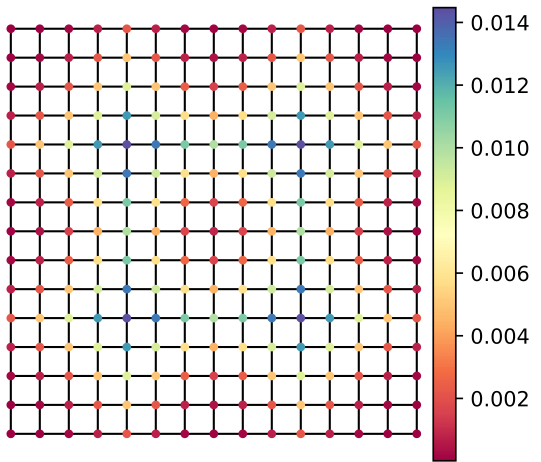}
  \caption{Synthetic map whose side length is \SI{1500}{m}. Axis represents the prior probability.}
  \label{fig:synth_unbr_map}
 \end{minipage}
\hfill
  \begin{minipage}{0.5\hsize}
   \centering\includegraphics[width=0.9\hsize]{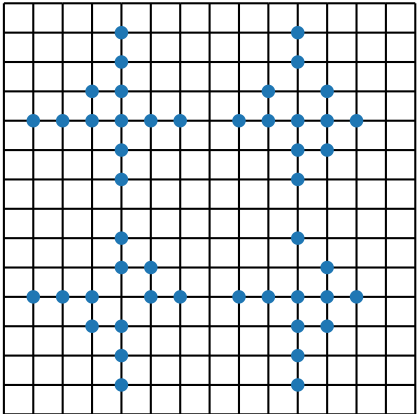}
  \caption{The example of the solution of the output range.}
  \label{fig:opt_example}
  \end{minipage}
\end{figure}

Here, we show examples of the optimization using the synthetic map. First, we explore the relationship between the number of nodes and the time required for the optimization (including initialization of $\mathcal{W}$). We use several lattices with different numbers of nodes(\fig~\ref{fig:square_graphs}). We use Python 3.7, an Ubuntu 15.10 OS, and 1 core of Intel core i7 6770k CPU with 64 GB of memory as the computational environment.
The results are shown in \fig~\ref{fig:time} and \fig~\ref{fig:PC_nodes}, where we can see that even when the number of nodes is large (e.g., $>5000$), the algorithm completes under $1$ minute and the PC improves by the optimization.
This time is acceptable because we can execute the algorithm to calculate future perturbations in advance.
As examples of the number of nodes, the two graphs in \fig~\ref{fig:maps} whose ranges are \SI{1000}{m} from the center contain $1,155$ and $168$ nodes, respectively. Even when a graph is quite large, by separating it into the small graphs such as those in \fig~\ref{fig:maps}, we can execute the algorithm in an acceptable time. Our implementation for the optimization is publicly available\footnote{https://github.com/tkgsn/GG-I}.

Next, we executed the algorithm using the synthetic map in \fig~\ref{fig:synth_unbr_map} under the assumption of the prior distribution. We assume that there are four places where the prior probability is high, as shown in \fig~\ref{fig:synth_unbr_map} and a user who follows this prior probability uses GEM with $\epsilon=0.01$ and an adversary has knowledge of the prior distribution. In this case, Q$^{loss}$ is \SI{328}{m} and PC is $0.9$ when we use $\mathcal{W}$ as all nodes.
A solution of the Algorithm~\ref{alg:opt} is as shown in \fig~\ref{fig:opt_example}.
By restricting output in the place where the prior probability is high, lower utility loss ($Q^{loss} =\SI{290}{m}$) and a higher tradeoff ($PC=0.98$) can be achieved.
The adversary infers that the pseudolocation is the true location, which means that the mechanism has effectively perturbed the true location.

\section{Experiments with Real-world Data}
\label{sec:experiments}

In this section, we show that GEM outperforms the baseline mechanism PLMG, which is the mechanism satisfying Geo-I, in terms of the tradeoff between utility and privacy on road networks of real-world maps.
\subsection{Comparison of GEM with PLMG}
\label{subsec:exp_gem_plmg}
\begin{figure}[t]
 \begin{minipage}{0.5\hsize}
   \centering\includegraphics[width=\hsize, height=\hsize]{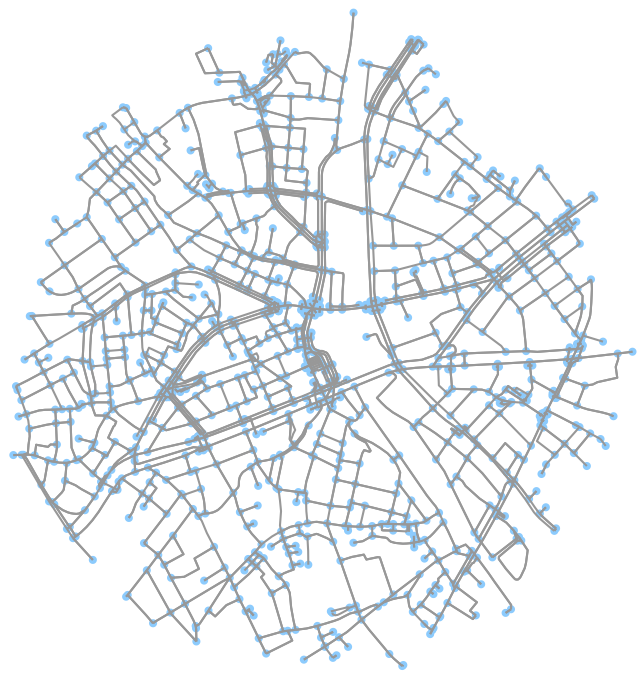}
 \end{minipage}
\hfill
 \begin{minipage}{0.5\hsize}
   \centering\includegraphics[height=\hsize]{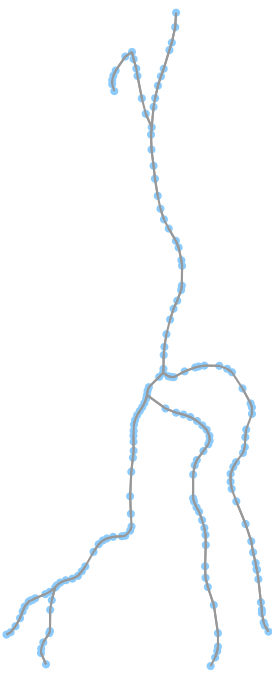}
 \end{minipage}
  \caption{On the left is a map of Tokyo, while the right shows a map of Akita.{\label{fig:maps}}}
\end{figure}

We evaluate the tradeoff of GEM based on the optimized range comparing with PLMG.
We use two kinds of maps~(\fig~\ref{fig:maps}) whose ranges are \SI{1000}{m} from the center, where points represent nodes. We assume that users are located in each node with the same probability. We use the output range of GEM obtained by Algorithm~\ref{alg:opt} according to this prior distribution.

\begin{figure}[t]
 \begin{minipage}{0.5\hsize}
  \centering\includegraphics[width=\hsize]{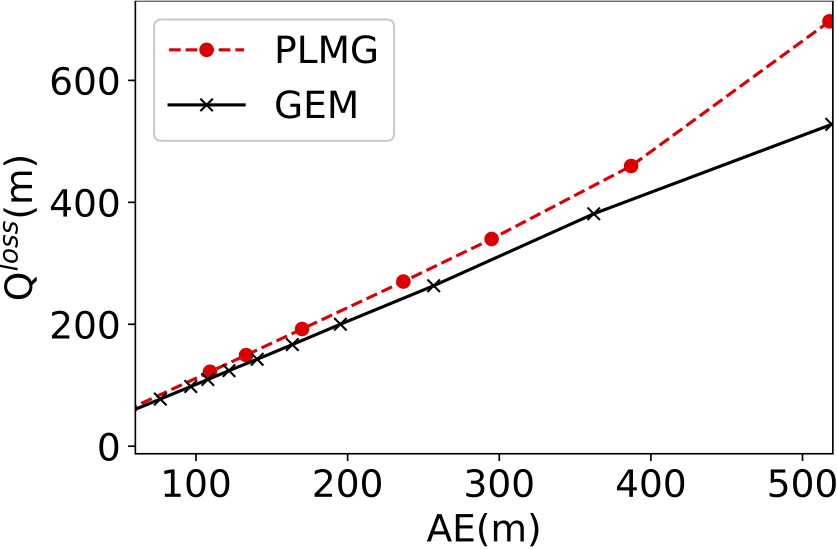}
 \end{minipage}
 \hfill
 \begin{minipage}{0.5\hsize}
  \centering\includegraphics[width=\hsize]{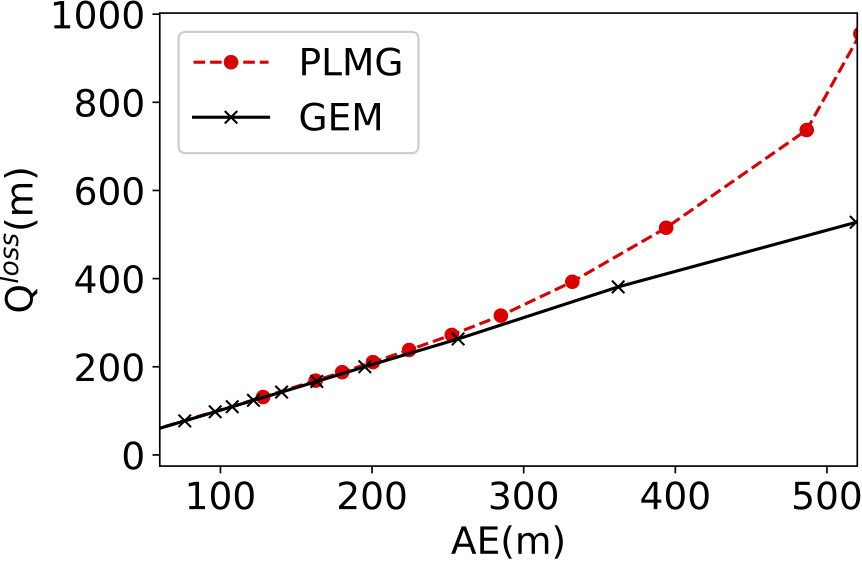}
 \end{minipage}
   \caption{Q$^{loss}$ comparison between PLMG and GEM on maps of Tokyo (left) and Akita (right) with respect to AE.{\label{fig:map_ae_sql}}}
\end{figure}

We compare Q$^{loss}$ of GEM with that of PLGM with respect to the same AE. Here, we assume an adversary who attacks with an optimal inference attack with the knowledge of the user, that is, the uniform distribution over the nodes. \fig~\ref{fig:map_ae_sql} shows that GEM outperforms PLMG in both maps w.r.t the trade-off between utility and privacy. Since GEM breaks the definition of Geo-I and tightly considers privacy of locations on road networks, GEM can output more useful locations than PLMG. Its performance advantage is greater on the Akita map because the difference between the Euclidean distance and the shortest distance is larger on that map than in on the Tokyo map.

\subsection{Evaluation of the Effectiveness of Optimization}
\fig~ \ref{fig:map_ae_sql} shows that the optimization works well. Here, we assume some prior knowledge and show the effectiveness of the optimization.
\subsubsection{Scenario}
\label{subsubsec:scenario}
\begin{figure}[t]
 \begin{minipage}{0.5\hsize}
  \centering\includegraphics[width=\hsize]{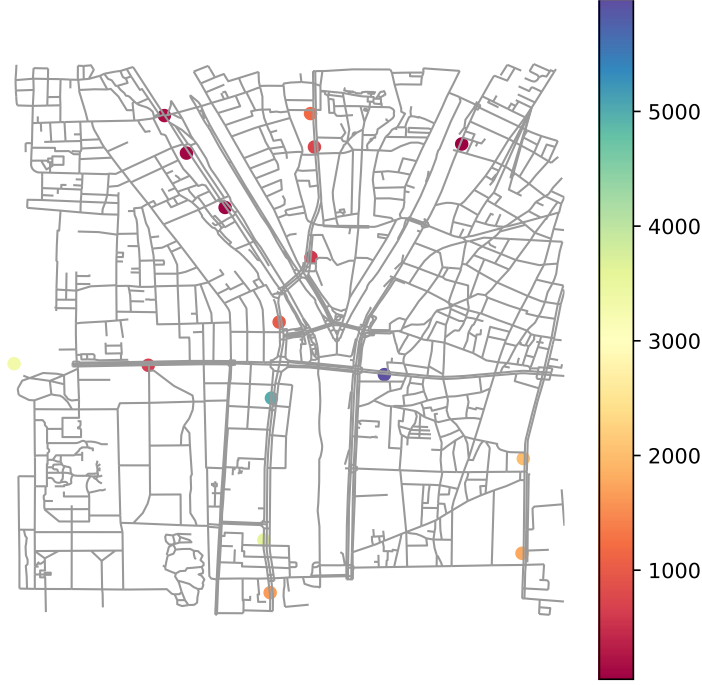}
     \caption{Each point represents a bus stop, and the y-axis represents the number of people who enter and exit buses at that stop.}
      \label{fig:kyoto_bus_stop}
 \end{minipage}
 \begin{minipage}{0.5\hsize}
  \centering\includegraphics[width=\hsize]{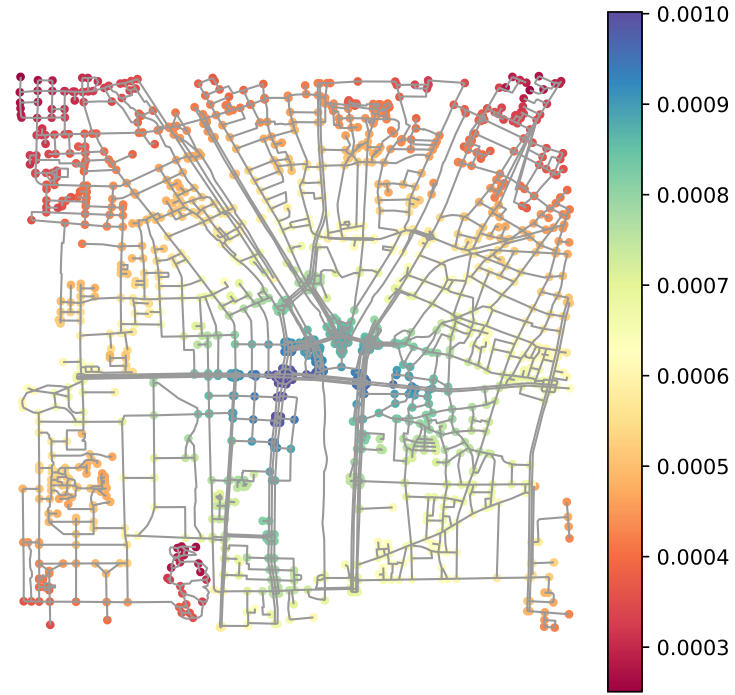}
     \caption{Prior distribution created from Kyoto Open Data.}
     \label{fig:kyoto_bus_prior}
 \end{minipage}
\end{figure}

\begin{figure}[t]
 \begin{minipage}{0.5\hsize}
  \centering\includegraphics[height=0.8\hsize]{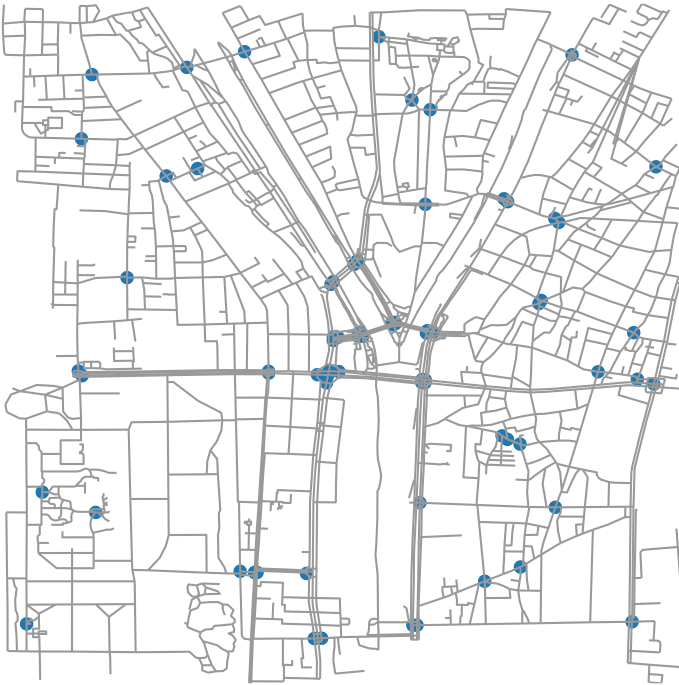}
     \caption{The solution for the objective function against the adversary with $\epsilon=0.01$.}
     \label{fig:kyoto_opt}
 \end{minipage}
 \hfill
 \begin{minipage}{0.5\hsize}
  \centering\includegraphics[width=\hsize]{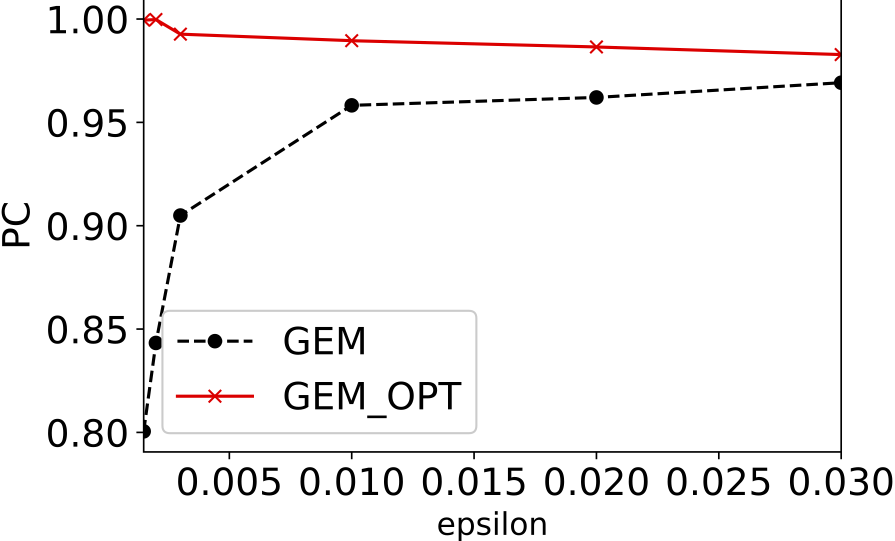}
     \caption{\mc\ with respect to $\epsilon$.}
      \label{fig:kyoto_epsilon_mc}
 \end{minipage}
\end{figure}

First, we show that the approximate solution for the proposed objective function effectively improves the tradeoff between utility and privacy. We use the following real-world scenario: a bus rider who uses LBSs. In other words, the user has a higher probability of being located near a bus stop. We create a prior distribution following this scenario by using a real-world dataset, Kyoto Open Data\footnote{https://data.city.kyoto.lg.jp/28}, which includes the number of people who enter and exit buses at each bus stop per day. \fig~\ref{fig:kyoto_bus_stop} shows the data, and \fig~\ref{fig:kyoto_bus_prior} represents the prior distribution made by distributing node probability based one the shortest distance from that node to a bus stop and the number of people who enter and exit buses at that bus stop. We assume that a user who follows this prior distribution uses an LBS with GEM and that an adversary knows the prior distribution. 
In this setting, we run Algorithm \ref{alg:opt} and obtain an approximate solution. Fig. \ref{fig:kyoto_opt} shows the example of an approximate solution. We can see that the nodes around the place with higher prior probability remain.

\subsubsection{Evaluation of Optimized Range}
First, we evaluate the PC of GEM with an optimized output range under the same $\epsilon$ as shown in \fig~\ref{fig:kyoto_epsilon_mc}. 
The result shows that a user can effectively perturb their true location for any realistic value of $\epsilon$ by using the optimized range. When the value of $\epsilon$ is small, the distribution of GEM has a gentle spread. In this case, the output of the mechanism does not contain useful information; thus, the adversary must use his/her prior knowledge, which results in a worse PC in the case of the baseline.
However, as these results show, by optimizing the output range according to the prior knowledge of the adversary, we can prevent this type of privacy leak.

\section{Related Works}
\label{sec:related}
\subsection{Cloaking}
Cloaking methods~\cite{duckham2005formal} obscure a true location by outputting an area instead of the true location. These methods are based on $k$-anonymity~\cite{spa-k-ano} which guarantees that at least $k$ users are in the same area, which prevents an attacker from inferring which user is querying the service provider. This privacy definition is practical, but there are some concerns~\cite{machanavajjhala2006diversity} regarding the rigorousness of the privacy guarantee because $k$-anonymity does not guarantee privacy against an adversary with some knowledge. If the adversary has peripheral knowledge regarding a user's location, such as range of the user's location, the obscured location can violate privacy. By considering the side knowledge of an adversary~\cite{xue2009location}, the privacy against that particular adversary can be guaranteed, but generally, protecting privacy against one type of adversary is insufficient. Additionally, introducing a cloaking method incurs additional costs for the service provider because the user sends an area rather than a location.

\subsection{Anonymization}
Anonymization methods~\cite{gedik2005location} separate a user's identifier from that user's location by assigning a pseudonym. Because tracking a single user pseudonym can leak privacy, the user must change the pseudonym periodically. Beresford et al.~\cite{mix-zone-orig} proposed a way to change pseudonyms using a place called mix zones. However, anonymization does not guarantee privacy because an adversary can sometimes identify a user by linking other information.

\subsection{Location Privacy on Road Networks}
To the best of our knowledge, this is the first study to propose a perturbation method with the differential privacy approach over road networks. However, several studies explored location privacy on road networks.\par
Tyagi et al.~\cite{tyagilocation} studied location privacy over road networks for VANET users and showed that no comprehensive privacy-preserving techniques or frameworks cover all privacy requirements or issues while still maintaining a desired location privacy level.\par
Wang et al.~\cite{wang2009privacy} and Wen et al.~\cite{Wen2018amethod} proposed a method of privacy protection for users who wish to receive location-based services while traveling over road networks. The authors used $k$-anonymity as the protection method and took advantage of the road network constraints.\par
A series of key features distinguish our solution from these studies: a) we use the differential privacy approach; consequently, our solution guarantees privacy protection against any attacker to some extent and b) we assume that no trusted server exists. We highlight these two points as advantages of our proposed method.

\subsection{State-of-the-Art Privacy Models}
Since Geo-I~\cite{geo-i} was published, many related applications have been proposed. To et al.~\cite{spatial-crowd} developed an online framework for a privacy-preserving spatial crowdsourcing service using Geo-I. Tong et al.~\cite{tong2017jointly} proposed a framework for a privacy-preserving ridesharing service based on Geo-I and the differential privacy approach. It may be possible to improve these applications by using GG-I instead of Geo-I. Additionally, Bordenabe et al.~\cite{opt-geo-i} proposed an optimized mechanism that satisfied Geo-I, and it may be possible to apply this method to GEM.\par
According to~\cite{geo-i}, using a mechanism satisfying Geo-I multiple times causes privacy degradation due to correlations in the data; this same scenario also applies to GG-I. This issue remains a difficult and intensely investigated problem in the field of differential privacy. Two kinds of approaches have been applied in attempts to solve this problem. The first is to develop a mechanism for multiple perturbations that satisfies existing notions, such as differential privacy and Geo-I~\cite{compo-theo,predic-dp}. Kairouz et al.~\cite{compo-theo} studied the composition theorem and proposed a mechanism that upgrades the privacy guarantee. Chatzikokolakis et al.~\cite{predic-dp} proposed a method of controlling privacy using Geo-I when the locations are correlated. The second approach is to propose a new privacy notion for correlated data~\cite{protec-locas,cao2018priste}. Xiao et al.~\cite{protec-locas} proposed $\delta$-location set privacy to protect each location in a trajectory when a moving user sends locations. Cao et al.~\cite{cao2018priste} proposed PriSTE, a framework for protecting spatiotemporal event privacy. We believe that these methods can also be applied to our work.

\section{Conclusion and Future Work}
In this paper, we proposed a new notion of location privacy on road networks, GG-I, based on differential privacy. GG-I provides a guarantee of the indistinguishability of a true location on road networks. We revealed that GG-I is a relaxed version of Geo-I, which is defined on the Euclidean plane. Our experiments showed that this relaxation allows a mechanism to output more useful locations with the same privacy level for LBSs that function over road networks. By introducing the notions of empirical privacy gain AE and utility loss Q$^{loss}$ in addition to indistinguishability $\epsilon$, we formalized the objective function and proposed an algorithm to find an approximate solution. We showed that this algorithm has an acceptable execution time and that even an approximate solution results in improved performance.

We represented a road network as a undirected graph; this means that our solution has no directionality even though one-way roads exist, which may degrade its utility.
In this paper, the target being protected is a location, but if additional information (such as which hospital the user is in) also needs to be protected, our proposed method does not work well: the hospital could be distinguished.
This problem can be solved by introducing another metric space that represents the targets to protect instead of the road network graph. Moreover, we need to consider the fact that multiple perturbations of correlated data, such as trajectory data, may degrade the level of protection even if the mechanism satisfies GG-I as in the case of Geo-I and differential privacy. This topic has been intensely studied, and we believe that the results can be applied to GG-I. 

\section{Acknowledgements}
This work is partially supported by the Japan Society for the
Promotion of Science (JSPS) Grant-in-Aid for Scientific Research (S)
No. 17H06099, (A) No. 18H04093, (C) No. 18K11314 and Early-Career Scientists
No. 19K20269.

\section{Appendix}
\setcounter{theorem}{0}
\setcounter{lemma}{0}

\begin{lemma}[Post-processing theorem of Geo-I.]
If a mechanism $M:\mathcal{X}\to\mathcal{Z}$ satisfies $\epsilon$-Geo-I, a post-processed mechanism $f\circ M$ also satisfies $\epsilon$-Geo-I for any function $f:\mathcal{Z}\to\mathcal{Z}^\prime$.
\end{lemma}

\begin{proof}
Given function $f:\mathcal{Z}\to\mathcal{Z}^\prime$, the following inequality holds for any two locations $x, x^\prime\in\mathcal{X}$ and $S\subseteq\mathcal{Z}^\prime$. We let $T$ denote $\{z\in \mathcal{Z}:f(z)\in S\}$; then, we have:
\begin{equation}
\begin{split}
\nonumber
\Pr(f(M(x))\in S) &= \Pr(M(x)\in T)\\
&\leq \mathrm{e}^{\epsilon d_e(x,x^\prime)} \Pr(M(x^\prime)\in T)\\
&= \mathrm{e}^{\epsilon d_e(x,x^\prime)} \Pr(f(M(x^\prime))\in S)\\
\end{split}
\end{equation}
This means that:
\begin{equation}
\nonumber
\log|\frac{\Pr(f(M(x))\in S)}{\Pr(f(M(x^\prime))\in S)}| \leq \epsilon d_e\\
\end{equation}
Q.E.D.
\end{proof}

\begin{theorem}
Given a graph $G=(V,E)$, GEM$_\epsilon$ satisfies $\epsilon$-GG-I.
\end{theorem}
\begin{proof}
We prove that the following inequality holds for any two locations on road networks $v,v^\prime\in V$ and $S\subseteq \mathcal{W}$:
\begin{equation}
\nonumber
\frac{\Pr(GEM(v)\in S)}{\Pr(GEM(v^\prime)\in S)}\leq \exp(\epsilon d_s(v,v'))
\end{equation}
The following inequality holds for any $S\subseteq\mathcal{W}$ and $v,v^\prime\in\ V$ from the triangle inequality:
\begin{equation}
\begin{split}
\nonumber
d_s(v,w)-d_s(v^\prime,w)&\leq d_s(v,v^\prime)
\end{split}
\end{equation}
Then, the left side of the inequality is transformed as follows:
\begin{equation}
\begin{split}
\nonumber
\frac{\Pr(GEM(v)\subseteq{S})}{\Pr(GEM(v^\prime)\subseteq{S})}=\frac{\alpha(v)\sum_{w\in S}\exp(-\epsilon d_v(v,w)/2)}{\alpha(v^\prime)\sum_{w\in S}\exp(-\epsilon d_{v^\prime}(v^\prime,w)/2)}\\
=\frac{\sum_{w\in\mathcal{W}}\exp(-\epsilon d_{v^\prime}(v^\prime,w)/2)}{\sum_{w\in\mathcal{W}}\exp(-\epsilon d_{v}(v,w)/2)}\frac{\sum_{w\in S}\exp(-\epsilon d_v(v,w)/2)}{\sum_{w\in S}\exp(-\epsilon d_{v^\prime}(v^\prime,w)/2)}\\
\leq  \exp(\epsilon d_s(v,v^\prime))
\end{split}
\end{equation}
Q.E.D.
\end{proof}
% you can choose not to have a title for an appendix
% if you want by leaving the argument blank
%\section{}
%Appendix two text goes here.

% use section* for acknowledgment
%\ifCLASSOPTIONcompsoc
  % The Computer Society usually uses the plural form
%  \section*{Acknowledgments}
%\else
  % regular IEEE prefers the singular form
%  \section*{Acknowledgment}
%\fi

% Can use something like this to put references on a page
% by themselves when using endfloat and the captionsoff option.
\ifCLASSOPTIONcaptionsoff
  \newpage
\fi

% trigger a \newpage just before the given reference
% number - used to balance the columns on the last page
% adjust value as needed - may need to be readjusted if
% the document is modified later
%\IEEEtriggeratref{8}
% The "triggered" command can be changed if desired:
%\IEEEtriggercmd{\enlargethispage{-5in}}

% references section

% can use a bibliography generated by BibTeX as a .bbl file
% BibTeX documentation can be easily obtained at:
% http://mirror.ctan.org/biblio/bibtex/contrib/doc/
% The IEEEtran BibTeX style support page is at:
% http://www.michaelshell.org/tex/ieeetran/bibtex/
%\bibliographystyle{IEEEtran}
% argument is your BibTeX string definitions and bibliography database(s)
%\bibliography{IEEEabrv,../bib/paper}
%
% <OR> manually copy in the resultant .bbl file
% set second argument of \begin to the number of references
% (used to reserve space for the reference number labels box)

\bibliographystyle{splncs04}
\bibliography{main}

%\begin{thebibliography}{1}

%\bibitem{IEEEhowto:kopka}
%H.~Kopka and P.~W. Daly, \emph{A Guide to \LaTeX}, 3rd~ed.\hskip 1em plus
%  0.5em minus 0.4em\relax Harlow, England: Addison-Wesley, 1999.

%\end{thebibliography}

% biography section
% 
% If you have an EPS/PDF photo (graphicx package needed) extra braces are
% needed around the contents of the optional argument to biography to prevent
% the LaTeX parser from getting confused when it sees the complicated
% \includegraphics command within an optional argument. (You could create
% your own custom macro containing the \includegraphics command to make things
% simpler here.)
%\begin{IEEEbiography}[{\includegraphics[width=1in,height=1.25in,clip,keepaspectratio]{mshell}}]{Michael Shell}
% or if you just want to reserve a space for a photo:

%\begin{IEEEbiography}{Michael Shell}
%Biography text here.
%\end{IEEEbiography}

% if you will not have a photo at all:
%\begin{IEEEbiographynophoto}{John Doe}
%Biography text here.
%\end{IEEEbiographynophoto}

% insert where needed to balance the two columns on the last page with
% biographies
%\newpage

%\begin{IEEEbiographynophoto}{Jane Doe}
%Biography text here.
%\end{IEEEbiographynophoto}

% You can push biographies down or up by placing
% a \vfill before or after them. The appropriate
% use of \vfill depends on what kind of text is
% on the last page and whether or not the columns
% are being equalized.

%\vfill

% Can be used to pull up biographies so that the bottom of the last one
% is flush with the other column.
%\enlargethispage{-5in}

% that's all folks

\end{document}